\documentstyle[aps,multicol,psfig]{revtex}

\newcommand{\PRB}[1]{Phys.\ Rev.\ B {\bf #1}}
\newcommand{\PRE}[1]{Phys.\ Rev.\ E {\bf #1}}
\newcommand{\PRL}[1]{Phys.\ Rev.\ Lett.\ {\bf #1}}

\newcommand{\PLA}[1]{Phys.\ Lett.\ A {\bf #1}}

\begin{document}

\draft

\title{
$1/f^{\alpha}$ fluctuations in a ricepile model
}
\author{
Shu-dong Zhang
} 
\address{Institute of Low Energy Nuclear Physics,
Beijing Normal University, Beijing 100875, China \\
Beijing Radiation Center, Beijing 100088, China \\
Institute of Theoretical Physics,
Beijing Normal University, Beijing 100875, China
}
\maketitle

\begin{abstract}
The temporal fluctuation of the average slope of a ricepile model is
investigated. It is found that the power spectrum $S(f)$ scales
as $1/f^{\alpha}$ with $\alpha\approx 1.3$ when grains of rice are added only
to one end of the pile. If grains are randomly added to the pile, the power
spectrum exhibits $1/f^2$ behavior. The profile fluctuations of the pile under
different driving mechanisms are also discussed.
\end{abstract}

\pacs{PACS numbers: 05.40.-a, 05.65.+b, 45.70.-n, 05.70.Ln}

\maketitle

\begin{multicols}{2}
The term {\em flick noise} refers to the phenomenon that a signal $s(t)$
fluctuates with a power spectrum $S(f)\propto 1/f^{\alpha}$ at low frequency.
Since the exponent $\alpha$ is often close to $1$, flick noise is also called
$1/f$ noise. The power spectrum of a signal is defined as the Fourier
transform of its auto-correlation function
$C(t_0,t)=\langle s(t_0)s(t)\rangle$, where $\langle \cdots \rangle$ denotes
ensemble average. Usually the signal $s(t)$ under study is stationary and its
auto-correlation function only depends on $t-t_0$. The auto-correlation
function can be alternatively calculated as
\begin{equation}
C(\tau)=\int_{-\infty}^{\infty}s(t)s(t+\tau)dt.
\end{equation}
If the signal is real-valued, its power spectrum is just
\begin{equation}
S(f)=\left| \widehat s(f)\right|^2
\label{eq_psd}
\end{equation}
according to Wiener-Khinchin Theorem.
Here $\widehat s(f)$ is the Fourier transform of the signal
\begin{equation}
\widehat s(f)=\int_{-\infty}^{\infty}s(t)e^{i2\pi ft}dt.
\end{equation}

In nature and laboratory, many physical systems show flick noise. For example,
flick noise appears in a variety of systems ranging from the light of quasars
\cite{Press1978} to water flows in rivers  \cite{Mandelbrot&Wallis1968}, music
and speech  \cite{Voss&Clarke1975}, and electrical measurement
\cite{Dutta&Horn1981,Weissman1988}. Despite its ubiquity, a universal
explanation of the flick noise is still lacking. Bak, Tang, and  Wiesenfeld
have proposed that self-organized criticality (SOC) \cite{Baketal1987} may be
the mechanism underlying flick noise. They also demonstrated their idea of SOC
with a cellular automaton model, the BTW sandpile model. However, the model
they proposed does not exhibit flick noise in its 1D and 2D versions
\cite{Jensenetalprb1989}. Especially the 1D BTW sandpile model does not even
exhibit SOC behavior. Experiments \cite{Fretteetalnature1996} on piles of rice
grains were done to investigate if real granular systems display SOC behaviors.
It was found that ricepiles with elongated rice grains exhibit SOC behaviors.
The key ingredient here is the heterogeneity of the system. Because elongated
rice grains could be packed in different ways \cite{Fretteetalnature1996},
the ricepile thus has many different metastable states. A 1D model of ricepile
was proposed by the same group of authors; avalanche distribution and
transit-time distributions were discussed \cite{Christensenetalprl1996}. The
model was later investigated by several other authors including us
\cite{Zhangpla1997,Bengrineetalictp1998,Amaral&Lauritsenpre1997}. Attentions
were paid to the avalanche dynamics and transit-time statistics of the model.
In this paper, we are concerned with temporal fluctuation of the
average slope of the system. We want to investigate if it has flick noise or
flick-like behavior.

The ricepile model is defined as follows: Consider a one-dimensional array of
sites $1$, $2$, $\cdots$, $L$. Each site contains an integer number $h_i$ of
rice grains. Here $h_i$ is called the local height of the ricepile at site
$i$. The system is initialized by setting $h_i=0$ for all sites. This means
the ricepile is to be built up from scratch. The system is driven by dropping
rice grains onto the pile. If one grain of rice is dropped at site $i$, then
the height of column $i$ will increase by $1$, $h_i\rightarrow h_i+1$. With
the dropping of rice grains, a ricepile is built up. The local slope of the
pile is defined as $z_i=h_i-h_{i+1}$. Whenever the local slope $z_i$ exceeds
certain threshold $z_i^c$ (specified in the following), site $i$ will topple
and one grain of rice will be transferred to its neighbor site on the right.
That is, $h_i\rightarrow h_i-1$ and $h_{i+1}\rightarrow h_{i+1}+1$. In this
model, rice grains are allowed to leave the pile only from the right end of
the pile, while the left end of the pile is closed. So the boundary condition
is kept as $h_0=h_1$ and $h_{L+1}=0$. The ricepile is said to be stable if no
local slope exceeds threshold value, that is, $z_i\leq z_i^c$ for all $i$.
Rice grains are dropped to the pile only when the pile is stable. In an
unstable state, all sites $i$ with $z_i>z_i^c$ topple in parallel. The
topplings of one or more sites is called an {\em avalanche} event, and the
size of an avalanche is defined as the number of topplings involved in the
avalanche event.

The values of $z_i^c$'s are essential to the definition of the model. As in
Ref. \cite{Zhangpla1997}, every time site $i$ topples, the threshold slope
$z_i^c$ will take a new value randomly chosen from $1$ to $r$ with equal
probability. Here $r$ is an integer no less than $1$. The
parameter $r$ reflects the level of medium disorder. The larger the $r$, the
higher the level of medium disorder of the system. If $r=1$ the model becomes
the $1$D BTW sandpile model. When $r=2$, the model reduces to the model
studied in reference \cite{Christensenetalprl1996}. In ref.
\cite{Zhangpla1997}, we have investigated the effect of disorder on the
universality of the avalanche size distribution and transit time distribution.

\begin{figure}
\centerline{\psfig{file=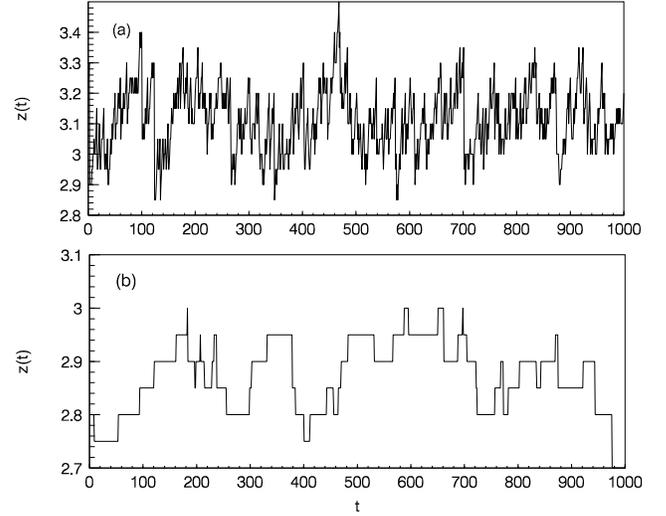,angle=270,width=1.0\columnwidth}}
\caption{
The time evolution of the average slope $z(t)$ of the ricepile after the
stationary state has been reached. The parameters used for this figure are
$L=20$, $r=4$. (a) rice grains are added to the pile at site $i=1$. (b) rice
grains are added randomly to the pile. 
}
\label{fig_zt}
\end{figure}

In the present work, we performed extensive numerical simulations on the
system evolution. Let
us first study the case where rice grains are added only to the left end of
the pile, i.e., only to the site $i=1$, which is in accordance with the
experiment setup \cite{Fretteetalnature1996}. We shall refer to this driving
mechanism as the {\em fixed driving}. When the stationary state is
reached, we record the average slope $z(t)=h_1(t)/L$ of the pile after every
avalanche. Here time $t$ is measured in the number of grained added to the
system. Thus we got a time series $z(t)$. Typical results about the
fluctuation of the slope $z(t)$ can be seen in Fig.\ref{fig_zt}.a. We
calculate the power spectrum of this time series according to
Eq.(\ref{eq_psd}). We find that for not too small systems the slope
fluctuation displays $1/f$-like behavior. In fact, we got a power spectrum
$S(f)$ scaling as $1/f^{\alpha}$ at low frequency, with the exponent
$\alpha=1.3\pm 0.1$.

We note the following points:

(a) the temporal behavior of the ricepile model is dramatically different from
the 1D BTW sandpile model. In the 1D BTW model, the system has a 
stable state \cite{Baketal1987}, which, once reached, cannot be altered by
subsequent droppings of sand grains. Thus the 1D BTW model does not display
SOC behaviors. The slope of the 1D sandpile model assumes constant values.
The power spectrum is thus of the form $S(f)=\delta (f)$. In words,  there
is only dc component in the power spectrum for the 1D BTW model.

(b) With the introduction of varying threshold $z_i^c$ into the
ricepile model the behavior of the system becomes much more rich. The
avalanche distribution follows power law \cite{Zhangpla1997}, which is an
important signature of SOC. Besides that, the temporal fluctuation of $z(t)$
has a power spectrum of flick type at low frequency.

\begin{figure}
\centerline{\psfig{file=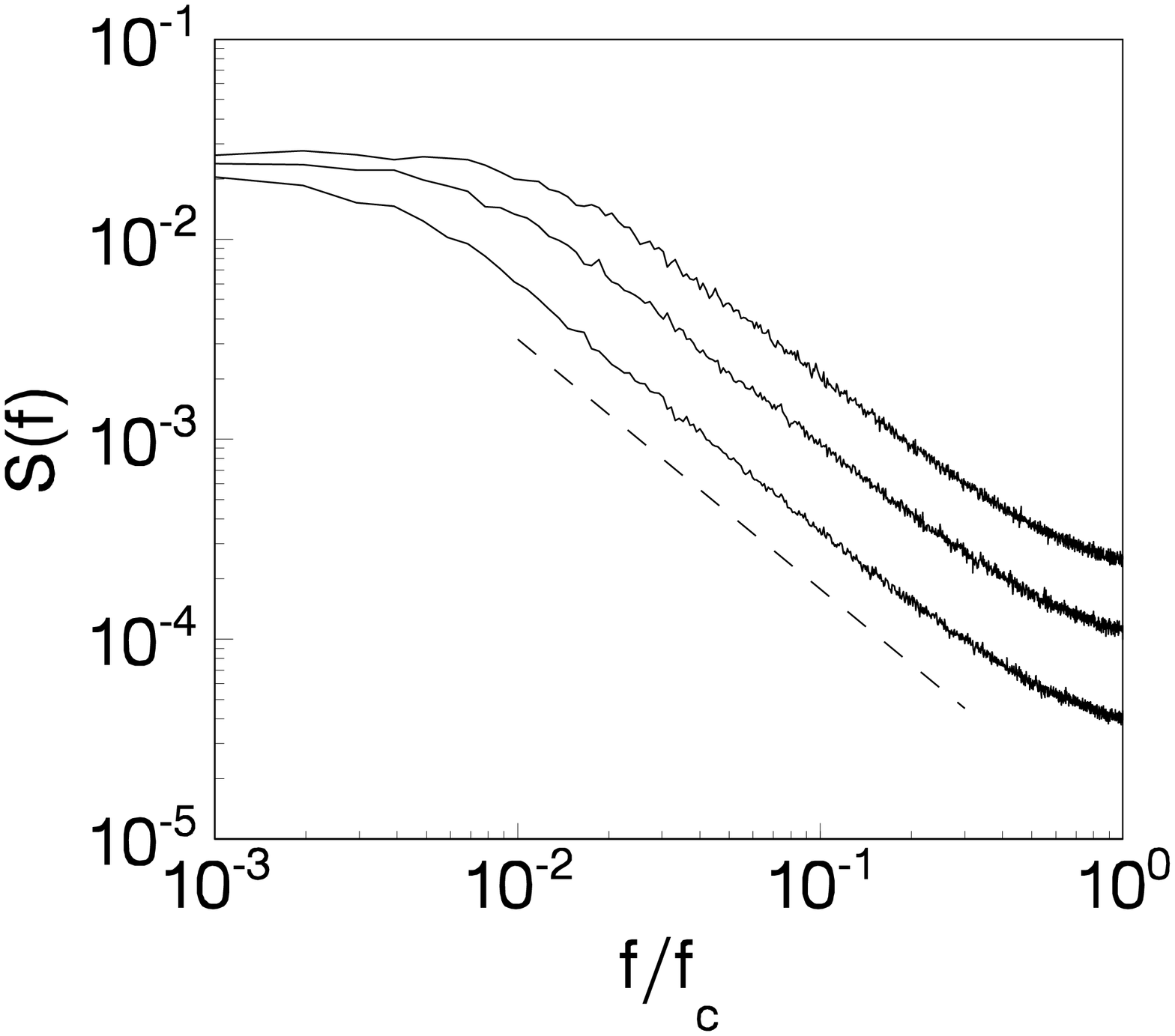,angle=270,width=1.0\columnwidth}}
\caption{
The power spectrum of $z(t)$ for different system sizes, with fixed 
driving and $r=2$. The
curves (from top to bottom) are for $L=40$, $60$, and $100$ respectively.
The dashed line is a curve for $y\propto x^{-1.25}$ for reference. The number
of data points for Fourier transformation is $N=2048$, and the results are
obtained by averaging over $500$ samples. $f_c\equiv 1/(2\Delta)$ is the
Nyquist critical frequency, where $\Delta$ is the time interval between two
successive points in the Fourier transformation.
}
\label{fig_r2lx}
\end{figure}

We notice that the power spectrum is flattened at low frequency for small
system. This is a size effect. It is believed that flick noise fluctuation
is closely related to the long range spatial correlation in the system. For
small system, the spatial correlation that can be built up is limited by the
system size, thus the long range temporal correlation required by flick
noise is truncated off at low frequency. Our numerical results verified the
above discussion. In Fig. \ref{fig_r2lx}, we show the power spectrum of $z(t)$
for different system sizes.  It is clear that when the
system size increases the $1/f^{\alpha}$ behaviors extends to lower and lower
frequency.

\begin{figure}
\centerline{\psfig{file=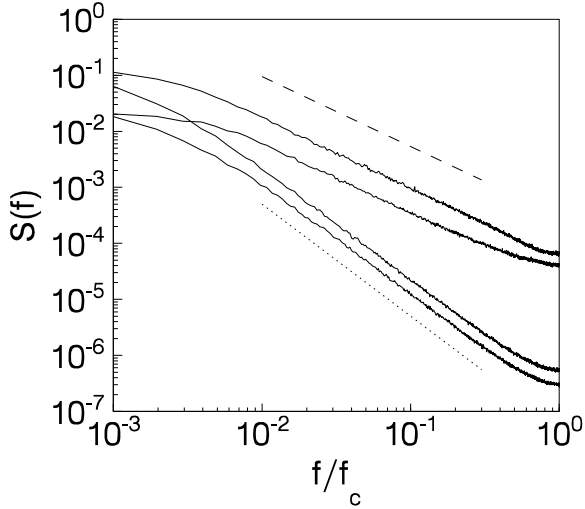,angle=270,width=1.0\columnwidth}}
\caption{
The power spectrum of the slope fluctuation for a ricepile with $L=100$. Two
groups of curves are shown in this figure, with different exponents $\alpha$.
For reference, the upper dashed line is a curve for $y\propto x^{-1.25}$,
while the lower dotted line shows a curve $y\propto x^{-2}$. In each group of
curves, the upper curve is for the case $r=4$, and the lower one is for $r=2$. 
}
\label{fig_l1hrx}
\end{figure}

We also investigated the effect of the level of disorder on the power
spectrum. This was done by simulating the system with different values of 
$r$. In Fig. \ref{fig_l1hrx} we show the power spectra for the system
with different $r$. It can be seen from this figure,  that
the power-law (straight in the log-log plot) part extends to lower frequency
for higher value of $r$. The effect can be understood by the following
discussion. When the level of disorder is increased, the
amplitude of the slope fluctuation also increases. Larger amplitude
fluctuation of $z(t)$ requires more grains to be added to the system.
Hence the period of this fluctuation increase, which gives rise to the
increase of low frequency components in the power spectrum. We have made
statistics on deviation of the ricepile slope from its average value in the
stationary state. We found that the greater the $r$, the greater the deviation. 
Note that the value of $r>1$ does not affect the exponent $\alpha$, as it does
not alter the universal avalanche exponent $\tau$ for the avalanche
distribution \cite{Zhangpla1997}.

It is also interesting to study the effect of driving mechanism on the
temporal behavior of the system. Now, instead of dropping rice grains to the
left end of the pile, we drop rice grains to randomly chosen sites of the
system. This way of dropping rice grains represents a different driving
mechanism, which we shall refer to as the {\em random driving}. Numerical 
simulations with this random driving were made and the
time series $z(t)$ was recorded. For this way of
driving, typical result about the fluctuation of $z(t)$ can be seen in
Fig. \ref{fig_zt}.b. It seems that $z(t)$ now fluctuates more regularly than 
it does for the fixed driving. We found that under random driving the
temporal fluctuation has a
trivial $1/f^2$ behavior as in the case of various previously studied
sandpiles \cite{Jensenetalprb1989}. In Fig. \ref{fig_l1hrx}, we compare the
power spectra of the two cases with different driving mechanisms. Two groups
of curves are shown in this figure, each with a different exponent
$\alpha$. For the fixed driving, we have $\alpha \approx 1.25$, while for the
random driving we have $\alpha \approx 2.0$. The distinction of the two
behaviors is quite clear. This shows that the driving mechanism, as an
integrated part of the model, has an important role in the temporal behaviors
of the system. Recent work on the continuous version of BTW sandpile
model\cite{Rios&Zhangprl1999}, and that on quasi-1D BTW
model\cite{Maslovetalprl1999} also showed that certain driving mechanisms are
necessary conditions for these models to display $1/f$ fluctuation for
the total amount of sand (or say, energy).

Here we present an heuristic discussion on why the exponent $\alpha$ is smaller
for fixed driving than for random driving. For fixed driving,
the height $h_1$ at site $i=1$ changes almost every dropping, and so is the 
average slope $z(t)$.
Thus the high frequency component of the $z(t)$ fluctuation has a
heavier weight in its power spectrum, and this will
make the exponent $\alpha$ smaller. While for the random driving, the
rice grains are dropped to the pile at random sites, the spatial correlations
previously built up can be easily destroyed, making the $z(t)$ behave more
or less
as an random walk. So the exponent $\alpha$ for this case shall be very close
to $2$. For random driving, each site has the same probability in
receiving a grain in every drop. For the fixed driving, however, only the
left end site receive grains, so there is in some sense breaking of symmetry,
which would lead to different scalings in the slope fluctuations.

\begin{figure}
\centerline{\psfig{file=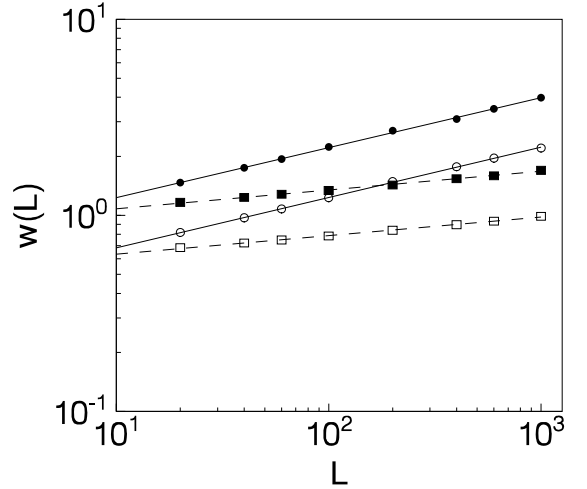,angle=270,width=1.0\columnwidth}}
\caption{
The profile width of the ricepile scales with the system size.
Data from numerical simulations are shown in open (for $r=2$)  or filled
( for $r=4$) symbols.
Solid lines are power law fit to the data for fixed driving,
with power law exponent $\chi =0.25\pm 0.01$,
and the dashed lines are for random driving, with $\chi =0.09\pm 0.01$.
For every run of numerical simulation, statistics were made over
at least $10^5$ grain dropping.
}
\label{fig_sigmaL}
\end{figure}

To see more about the effect of driving mechanism, it is helpful to investigate 
the profile fluctuations
of the pile under different drivings. Because avalanches
change the surface of the ricepile, the pile fluctuates around an
average profile, and the size of the fluctuations characterize the active zone
of the ricepile surface. Let the standard deviation of height at site $i$ be
$\sigma _h(i,L)=\sqrt{\langle h_i^2\rangle -\langle h_i\rangle ^2}$. Here 
$\langle \cdots \rangle$ 
represents average over time. Then we calculate the profile width of the 
ricepile,
$w={1\over L}\sum _i \sigma _h(i,L)$,
which is a function
of the system size. In Fig. \ref{fig_sigmaL}, we show the profile width of
the ricepile for different $r$ and different driving mechanisms. It can be
seen that $w$ scales with $L$ as
$w\propto L^{\chi }$, and that $\chi =0.25\pm 0.01$ for
fixed driving, $\chi =0.09\pm 0.01$ for random driving.
For given driving mechanism, $w$ increases with increasing $r$. As we stated
in Ref. \cite{Zhangpla1997},
the parameter $r$ reflects the level of medium disorder in the rice pile.
Although the level of disorder does not change the scaling exponent $\chi $,
it does affect the amplitude of fluctuations. For greater $r$, the profile
width is larger. It can be seen in the figure, that the data points
for $r=4$ are above that for $r=2$, for given driving mechanism.
A recent experiment showed that piles of polished rice grains
has a smaller profile width than unpolished ones\cite{Federetalprl1999},
which have higher level of medium disorder.

In conclusion, we have studied the temporal fluctuation of the slope of
ricepile model in its critical stationary state. The power spectrum of this
fluctuation is closely related to the driving mechanism. When
the rice grains are dropped to the left end of the pile, the slope fluctuates
with a flick-type power spectrum, with the exponent $\alpha=1.3\pm 0.1$. When
the driving mechanism is changed to random driving, the model displays $1/f^2$
behaviors. Greater system size and higher level of disorder will extend the 
frequency range where the power spectrum has the form $1/f^{\alpha}$.

The author thanks Prof. Vespignani for helpful discussions.
This work was supported by the National Natural Science Foundation of China
under Grant No. 19705002, and the Research Fund for the Doctoral Program of
Higher Education (RFDP).

\end{multicols}
\end{document}